\documentclass[aps,prb,twocolumn,showpacs]{revtex4}
\usepackage{amssymb}
\usepackage{graphicx}

\begin{document}

\title{Magnetization relaxation and collective vortex pinning in Fe-based superconductor SmFeAsO$_{0.9}$F$_{0.1}$ }

\author{Huan Yang, Cong Ren, Lei Shan, and Hai-Hu Wen}\email{hhwen@aphy.iphy.ac.cn }

\affiliation{National Laboratory for Superconductivity, Institute of
Physics and Beijing National Laboratory for Condensed Matter
Physics, Chinese Academy of Sciences, P. O. Box 603, Beijing 100190,
People's Republic of China}

\begin{abstract}
By measuring the dynamic and traditional magnetization relaxations
we investigate the vortex dynamics of the recently discovered
superconductor SmFeAsO$_{0.9}$F$_{0.1}$ with $T_\mathrm{c} = 55\;$K.
It is found that the relaxation rate is rather large reflecting a
small characteristic pinning energy. Moreover it shows a weak
temperature dependence in wide temperature region, which resembles
the behavior of the cuprate superconductors. Combining with the
resistivity data under different magnetic fields, a vortex phase
diagram is obtained. Our results strongly suggest that the model of
collective vortex pinning applies to this superconductor very well.

\end{abstract} \pacs{74.25.Qt, 74.25.Ha, 74.70.Dd}
\maketitle

Since the discovery of superconductivity at $T_\mathrm{c}=26\;$K
\cite{LaOFFeAs} in LaFeAsO$_{1-x}$F$_x$, the iron based layered
superconductors have exposed an interesting research area on
superconductivity. This family of superconductors,
LnFeAsO$_{1-x}$F$_x$, exhibit quite high critical temperatures with
the maximum $T_\mathrm{c}=55\;$K for Ln = Sm \cite{SmF_RZA} in
electron doped region, as well as $25\;$K in the hole doped case
La$_{1-x}$Sr$_x$FeAsO \cite{LaSr_Wen}. Lots of experimental and
theoretical works on the physical properties were accomplished.
Measurements under high magnetic fields reveal that the iron based
superconductors have very high upper critical
fields\cite{LaF_Hunte,Senatore,JiaY}, which indicates encouraging
potential applications. It was pointed out that these
superconductors exhibit multiband feature
\cite{LaF_Zhu,LaF_Ren,LaF_Hunte} as well as the unconventional
pairing symmetry \cite{LaF_Ren,LaF_Shan,LaF_Mu}. Moreover this
system has a layered structure with the conducting FeAs layers being
responsible for the superconductivity, and the LnO layers behave as
the charge reservoirs, all these look very similar to the case of
cuprates. For cuprate superconductors, due to the high anisotropy,
short coherence length and high operation temperature, the vortex
motion and fluctuation are quite strong. This leads to a small
characteristic pinning energy, and the single vortex or vortex
bundles are pinned collectively by many small pinning
centers\cite{Blatter}. Therefore it is curious to know whether the
vortex properties and phase diagram of the cuprate and FeAs-based
superconductors are similar to each other or not. The magnetization
relaxation has been proved to be a very effective way to investigate
the vortex dynamics\cite{Yeshurun,Brandt}. In this paper, we report
a detailed study on the vortex dynamics of SmFeAsO$_{0.9}$F$_{0.1}$
polycrystalline samples.

\begin{figure}
\includegraphics[width=8cm]{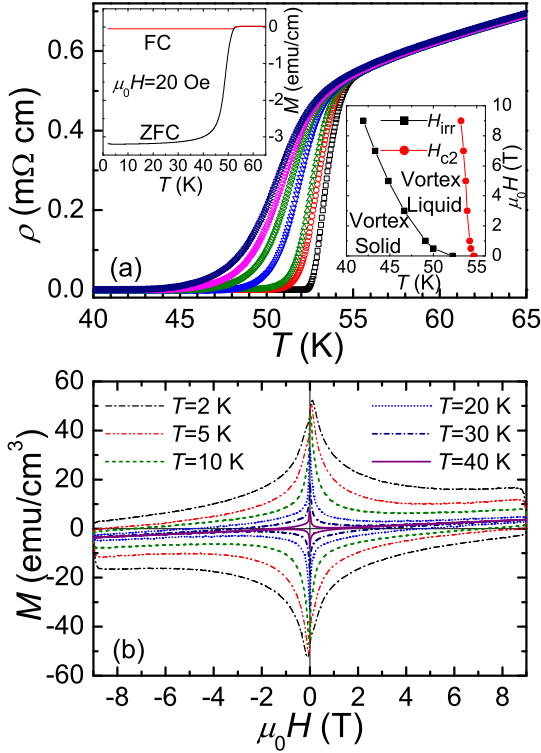}
\caption{(Color online) (a) Temperature dependence of resistivity at
various fields of 0, 0.5, 1, 3, 5, 7, 9$\;$T. The upper inset in (a)
shows the temperature dependence of the diamagnetic moment measured
in the ZFC and FC processes at a field of 20$\;$Oe, while the inset
below shows the vortex phase diagram (see text). (b) Magnetization
hysteresis loops $M$ vs $H$ at different temperatures of the same
sample. } \label{fig1}
\end{figure}

The SmFeAsO$_{0.9}$F$_{0.1}$ samples used in our measurements were
grown by the high pressure synthesis method\cite{SmF_RZA}. The
sample in this work was first cut into a rectangular shape with the
dimensions of
$4.20\;\mathrm{mm}\times1.60\;\mathrm{mm}\times1.08\;\mathrm{mm}$
for the resistance measurement, and it was further shaped into a bar
shape with $2.68\;\mathrm{mm}$ in length (width and thickness
unchanged) for the magnetic measurement. The measurements were
carried out with a physical property measurement system (PPMS,
Quantum Design) with the magnetic field up to 9$\;$T. The magnetic
field sweeping rate can be varied from 0.5 to $600\;$Oe/s. The
magnetic measurements were made by the sensitive vibrating sample
magnetometer (VSM) at the vibrating frequency of $40\;$Hz with the
resolution better than $1\times10^{-6}\;$emu. The advantage of this
technique is that the data acquisition is very fast with a quite
good resolution for magnetization.

In the upper inset of Fig.~\ref{fig1} (a), we show the diamagnetic
transition of the sample measured in the field-cooled (FC) and
zero-field-cooled (ZFC) processes. The ZFC curve shows perfect
diamagnetism in the low temperature region when taking the
demagnetization factor into account. In Fig.~\ref{fig1} (a), we show
the $\rho$-$T$ curves at different magnetic fields. The onset
transition temperature taken with a criterion of
$99\%\rho_\mathrm{n}$ at zero field is about $55.2\;$K, while the
zero-resistance temperature is about $52.2\;$K. Both the narrow
magnetic and resistive transition widthes indicate good quality of
the polycrystalline sample. The coherence length of the iron based
superconductor is supposed to be larger than that of cuprates, and
the sample grown by the high pressure synthesis method is very
dense, so it warrants a further detailed investigation on vortex
dynamics. From the $\rho-T$ curves, we obtained the phase diagram as
depicted in the inset of Fig.~\ref{fig1} (a) (the  $H_{c2}(T)$ was
obtained by using the criterion of $95\%\rho_\mathrm{n}$). The ratio
between the irreversibility line $H_\mathrm{irr}(T)$ and the upper
critical field $H_\mathrm{c2}(T)$ is close to that of
YBa$_2$Cu$_3$O$_{7-\delta}$ (YBCO), but much larger than that of the
more anisotropic Bi-based cuprate system. The calculated $H_{c2}(0)$
at zero temperature is $312\pm26\;$T by using the
Werthamer-Helfand-Hohenberg (WHH) formula \cite{WHH}
$H_\mathrm{c2}(0)=-0.69\mathrm{d}H_\mathrm{c2}(T)/\mathrm{d}T|_{T_\mathrm{c}}T_\mathrm{c}$
roughly, and $444\pm16\;$T by fitting $H_{c2}(T)$ with the
expression
$H_\mathrm{c2}(T)=H_\mathrm{c2}(0)\left[1-(T/T_\mathrm{c})^2\right]/\left[1+(T/T_\mathrm{c})^2\right]$
based on the Ginzburg-Landau theory. In Fig.~\ref{fig1} (b) we show
the magnetization hysteresis loops (MHL) measured at different
temperatures from 2$\;$K to 50$\;$K. The symmetric curves indicate
that the bulk current instead of the surface shielding current
dominates in the sample. It is remarkable that the superconducting
MHL can still be measured at temperatures very close to
$T_\mathrm{c}$, with only a weak magnetic background. This indicates
that the sample contains negligible magnetic impurities. Based on
the Bean critical state model \cite{Bean}, the superconducting
current density $j\propto \Delta M$, where $\Delta M=M^--M^+$, and
$M^+$ ($M^-$) is the magnetization associated with increasing
(decreasing) field.

\begin{figure}
\includegraphics[width=8cm]{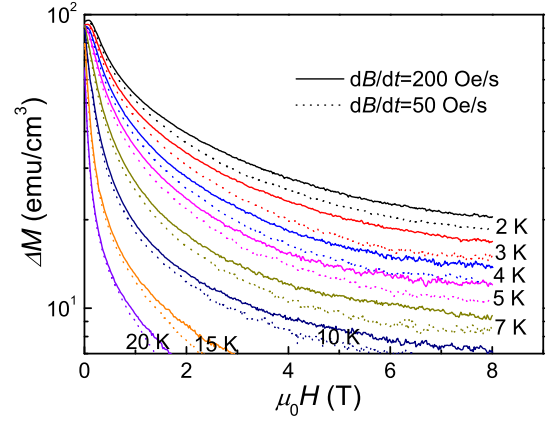}
\caption{(Color online) Field dependence of $\Delta M$ with
different field sweeping rates of 200$\;$Oe/s and 50$\;$Oe/s. }
\label{fig2}
\end{figure}

\begin{figure}
\includegraphics[width=8cm]{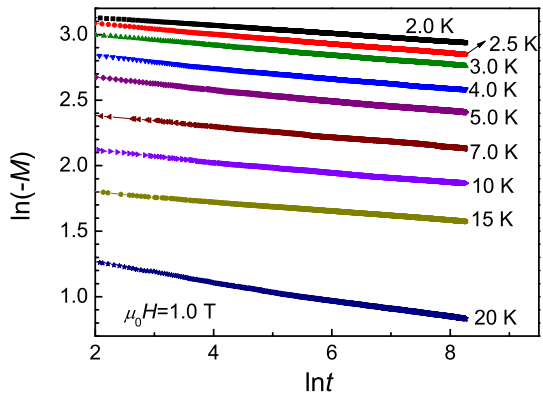}
\caption{(Color online) Log-log plot of magnetization $-M$
(emu/cm$^3$) vs. time $t$ (sec) at various temperatures at
$\mu_0H=1.0\;$T. } \label{fig3}
\end{figure}

In a type-II superconductor, the vortices normally move through
thermal activation over the effective pinning barrier $U(j,T,B)$
with an average velocity $\bar v = v_0
\exp[-U(j,T,B)/k_\mathrm{B}T]$. Here $v_0$ is the attempt hopping
velocity. The effective pinning barrier or the activation energy can
be written as\cite{Malozemoff}
\begin{equation}
U(j,T,B)=\frac{U_\mathrm{c}(T,B)}{\mu_(T,B)}
\left[\left(\frac{j_\mathrm{c}(T,B)}{j(T,B)}\right)^{\mu(T,B)}-1\right],\label{UJ}
\end{equation}
where $U_\mathrm{c}$ is the characteristic pinning energy, $\mu$ is
the glassy exponent\cite{Fisher,Vinokur}, and $j_\mathrm{c}$ is the
critical current density. The dissipation is associated with an
electric field $E = \bar v B = v_0B\exp[-U(j,T,B)/k_\mathrm{B}T]$,
where $B$ is the local magnetic induction. As proposed by Schnack
\textit{et al.} \cite{DMRM1} and Jirsa \textit{et al.} \cite{DMRM2},
the magnetization-relaxation measurements can be measured with
different magnetic sweeping rates $\mathrm{d}B/\mathrm{d}t$ which
was called as the dynamic magnetic relaxation measurements. The term
``dynamic'' here originates from a comparison with the traditional
relaxation method, that is to measure the time dependence of the
magnetization after the field sweeping is stopped. The corresponding
magnetization relaxation rate is defined as $Q\equiv\mathrm{d}\ln
j/\mathrm{d}\ln\left(\mathrm{d}B/\mathrm{d}t\right)=\mathrm{d}\ln
\Delta M/\mathrm{d}\ln\left(\mathrm{d}B/\mathrm{d}t\right)$. As
shown in Fig.~\ref{fig2}, $\Delta M$ is obviously different from
each other with different field sweeping rates 200$\;$Oe/s and
50$\;$Oe/s: a faster sweeping rate corresponds to a higher instant
current density. A surprising observation here is that the gap
between $\Delta M$ measured at 200$\;$Oe/s and 50$\;$Oe/s can be
easily distinguished even at a low temperature, this indicates a
relatively large vortex creep rate, or as called as giant vortex
creep in the cuprate superconductors. The calculated $Q$ from
Fig.~\ref{fig2} is as large as 5\% at $1\;$T. This value is similar
to the one in cuprate superconductors, e.g. 4\% in YBCO
\cite{Yeshurun,YBCOWen}, but one order of magnitude larger than that
in MgB$_2$ at such a small field\cite{MgB2Wen}. In order to check
the relaxation rate derived from the dynamic relaxation method, we
also did the traditional magnetization-relaxation measurements on
this sample. In this case, the normalized magnetization relaxation
rate $S$ is defined as $\mathrm{d}\ln (-M)/\mathrm{d}\ln t$. The
time ($t$) dependence of the non-equilibrium magnetization ($M$) are
shown in Fig.~\ref{fig3} on a log-log plot. It shows that $\ln(-M)$
decays with time in a logarithmic way at 1$\;$T, which is actually
expected by the model of thermally activated flux motion.

\begin{figure}
\includegraphics[width=8cm]{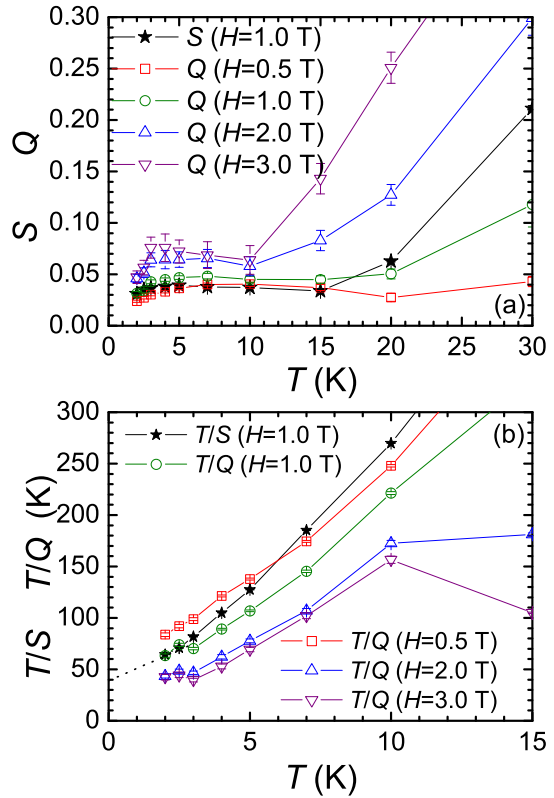}
\caption{(Color online) (a) Temperature dependence of the dynamic
relaxation rate $Q$ at different magnetic fields and the normalized
relaxation rate $S$ at 1 T. (b) Temperature dependence of $T/S$ or
$T/Q$ at various fields. } \label{fig4}
\end{figure}

In Fig.~\ref{fig4} (a), we present the temperature dependence of the
two relaxation rates $Q$ and $S$, and they exhibit the similar
temperature dependence. Obviously, there is a plateau in the
intermediate temperature region for each curve, and the region of
the plateau increases with decreasing the magnetic field. This
plateau cannot be understood within the picture of single vortex
creep with the rigid hopping length as predicted by the Anderson-Kim
model. Exactly the same behavior was observed in YBCO \cite{YBCO_cp}
and was attributed to the vortex collective pinning in the cuprate
superconductor. However, this is in contrast to the data in MgB$_2$
where the relaxation rate increases linearly with temperature,
indicating a thermally activated hopping of vortices with rigid
length.\cite{MgB2Wen} The general feature of relaxation rate shown
in Fig.\ref{fig4}(a) is very similar to the case of cuprate
superconductors, especially in YBCO. \cite{Griessen}  In our
experiment, the lowest temperature was only down to 2$\;$K. So it is
difficult to know whether the quantum tunneling of vortices occurs
at a much lower temperature.
To get a comprehensive understanding about the vortex motion in the
intermediate temperature region, we use the following expression to
calculate the characteristic pinning energy \cite{Tl_Wen}
\begin{equation}
\frac{T}{Q(T,B)}(\mathrm{or\ }
\frac{T}{S(T,B)})=\frac{U_\mathrm{c}(T,B)}{k_\mathrm{B}}+\mu(T,B)
CT,\label{TQ}
\end{equation}
where $C\simeq\ln(2v_0B/l\mathrm{d}H/\mathrm{d}t)$ is a parameter
which is weakly temperature dependent. In the low temperature region
(below 3$\;$K), the relaxation rate has a clear tendency to drop
with temperature. This can be understood based on the picture of
vortex collective pinning: According to Eq.~\ref{TQ}, when $\mu CT$
becomes smaller than $U_\mathrm{c}(T,B)$, we have $T/Q \approx
U_\mathrm{c}(T,B)/k_\mathrm{B}$ and $Q$ rises almost linearly with
$T$. In the intermediate temperature region, $\mu CT$ is getting
gradually larger than $U_\mathrm{c}(T,B)$, the relaxation rate $Q$
is thus determined by the balance between them. In
Fig.~\ref{fig4}(b), we show the temperature dependence of $T/Q$ or
$T/S$ vs. $T$. By extrapolating the curve $T/S$ or $T/Q$ down to
zero temperature, one can get the value of $U_\mathrm{c}(0)$. The
value of $U_\mathrm{c}(0)/k_\mathrm{B}$ at $1\;$T calculated from
Fig.~\ref{fig4} is about $40\;$K, which is a very small value. The
$U_\mathrm{c}(0)$ is about $100\sim400\;$K in YBCO thin films
\cite{YBCOWen}, but beyond $3000\;$K in MgB$_2$ \cite{JinH}.
Meanwhile, the parameter $C$ in Eq.~\ref{TQ} can be determined from
the curve $-\mathrm{d}\ln j/\mathrm{d}T$ vs. $Q/T$. \cite{Tl_Wen}
And here we find $C=7.3\pm0.3$. The slope of $T/Q$ vs $T$ that gives
the value of $\mu C$ in low temperature region at $1\;$T is about
11.8, so we get the parameter $\mu=1.62$. This value is close to
$\mu=3/2$ predicted for collective pinning of small bundles.

\begin{figure}
\includegraphics[width=8cm]{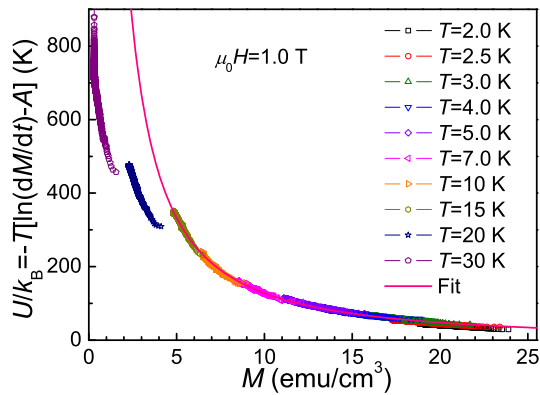}
\caption{(Color online) Magnetization dependence of the scaled
activation energy $U$ obtained by following the Maley's technique
with parameter $A=8$. The line is a fitting result by Eq.~\ref{UJ}
with $U_\mathrm{c}/k_\mathrm{B}=39.7\;$K and $\mu=1.11$.}
\label{fig5}
\end{figure}

In the following we analyze the $U(j)$ relation by Maley's
method\cite{Maley}, which proposed a general equation
$U/k_\mathrm{B}=-T[\ln\left(\mathrm{d}M/\mathrm{d}t\right)-A]$ to
scale the data measured at different temperatures, where $A$ is a
time independent constant associated with the average hopping
velocity. In Fig.~\ref{fig5}, we present the correlation between the
scaled activation energy $U$ and $M$. Obviously, at the temperatures
below 15$\;$K, all the curves can be scaled together, but the
scaling fails at temperatures above 20$\;$K. The reason may be that
the original Maley's method is valid only in low temperature regime
(below $T_\mathrm{c}/3$) where the explicit temperature dependence
of $U_\mathrm{c}(T)$ is weak\cite{Maley}. It should be noted that
the relaxation rate at $1\;$T has a sudden increase which starts at
about 20$\;$K after the plateau, this may correspond to a crossover
between different vortex creep regimes. A recent work by the
magneto-optical imaging technology showed that the local current
density exhibited different temperature dependence above and below
20$\;$K,\cite{Yamamoto} which may be caused by the similar reason.
According to Eq.~\ref{UJ}, we can fit the activation energy data by
the thermally activated vortex motion theory. The $U-M$ curves at
the temperature below 15$\;$K can be well fitted (as shown by the
solid line in Fig.~\ref{fig5}). The fitting parameter
$U_\mathrm{c}/k_\mathrm{B}$ is 39.7$\;$K which is close to $40\;$K
mentioned above. From the fit we get $\mu=1.11$, so the value
$\mu=3/2$ for small bundle is just in between the two calculated
values of $\mu$ obtained by different methods in this work. Our work
strongly suggest that the collective pinning model is applicable in
this kind of superconductors. This is consistent with a recent work
on the vortex dynamics of the polycrystalline
NdFeAsO$_{0.9}$F$_{0.1}$ sample measured by local magneto-optical
imaging technique.\cite{Prozorov} From the $U-M$ relation shown in
Fig.~\ref{fig5}, one can also see that the activation energy
increases sharply with decreasing the current density. In the
resistive measurements, the current density is usually very small
(about $0.3\;\mathrm{A/cm}^2$ in this work), so the activation
energy obtained from the Arrhenius plot of $R$-$T$ curves may reach
a quite large value.\cite{Dou}

In conclusion, dynamic and traditional magnetization relaxation have
been measured on SmFeAsO$_{0.9}$F$_{0.1}$ samples in wide
temperature and magnetic field regions. The model of collective
vortex pinning seems to work in understanding the vortex properties
of this material. The relaxation rate is quite large which is
comparable with the cuprate superconductors and much larger than the
value in MgB$_2$. The characteristic pinning energy is only about
$40\;$K, but the vortex (or vortex bundles) are pined collectively
by many weak pinning centers.

\begin{acknowledgments}
We are grateful to Prof. Zhongxian Zhao and Dr. Zhian Ren for
providing us the high quality SmFeAsO$_{0.9}$F$_{0.1}$ samples made
by high pressure technique. This work is supported by the Natural
Science Foundation of China, the Ministry of Science and Technology
of China (973 project No: 2006CB601000, 2006CB921802), and Chinese
Academy of Sciences (Project ITSNEM).
\end{acknowledgments}


\begin{thebibliography}{00}

\bibitem{LaOFFeAs} Y. Kamihara, T. Watanabe, M. Hirano, and H. Hosono, J. Am. Chem. Soc. \textbf{130}, 3296 (2008).

\bibitem{SmF_RZA} Z. A. Ren, W. Lu, J. Yang, W. Yi, X. L. Shen, C. Z. Li, G. C. Che,
 X. L. Dong, L. L. Sun, F. Zhou, and Z. X. Zhao, Chin. Phys. Lett. \textbf{25}, 2215
(2008).

\bibitem{LaSr_Wen} H. H. Wen, G. Mu, L. Fang, H. Yang, and X. Zhu, Europhys. Lett. \textbf{82}, 17009
(2008).

\bibitem{LaF_Hunte} F. Hunte, J. Jaroszynski, A. Gurevich, D. C. Larbalestier, R. Jin, A. S. Sefat,
 M. A. McGuire, B. C. Sales, D. K. Christen, and D. Mandrus, Nature (London) \textbf{453}, 903 (2008).

\bibitem{Senatore} C. Senatore, R. Fl\"{u}kiger, M. Cantoni, G. Wu, R. H. Liu, and X. H. Chen, Phys. Rev. B \textbf{78} 054514 (2008).

\bibitem{JiaY} Y. Jia, P. Cheng, L. Fang, H. Q. Luo, H. Yang, C. Ren, L. Shan, C. Z. Gu, and H. H. Wen, Appl. Phys. Lett. \textbf{93}, 032503
(2008).

\bibitem{LaF_Zhu} X. Y. Zhu, H. Yang, L. Fang, G. Mu, and H. H. Wen, Supercond. Sci. Technol.
\textbf{21}, 105001(2008).

\bibitem{LaF_Ren} C. Ren,  Z. S. Wang, H. Yang, X. Y. Zhu, L. Fang, G. Mu, L. Shan, and H. H. Wen, arXiv:0804.1726.

\bibitem{LaF_Shan} L. Shan, Y. L. Wang, X. Y. Zhu, G. Mu, L. Fang, and H. H. Wen, Europhys. Lett. \textbf{83}
57004 (2008).

\bibitem{LaF_Mu} G. Mu, X. Y. Zhu, L. Fang, L. Shan, C. Ren, and H. H. Wen, Chin. Phys. Lett. \textbf{25}, 2221
(2008).

\bibitem{Blatter} G. Blatter, M. V. Feigel'man, V. B. Geshkenbein, A. I. Larkin, and V. M. Vinokur, Rev. Mod. Phys. \textbf{66},
1125 (1994).

\bibitem{Yeshurun} Y. Yeshurun, A. P. Malozemoff, and A. Shaulov,
Rev. Mod. Phys. \textbf{68}, 911 (1996).


\bibitem{Brandt} E. H. Brandt, Rep. Pog. Phys. \textbf{58}, 1465 (1996).


\bibitem{WHH} N. R. Werthamer, E. Helfand, and P. C. Hohenberg, Phys. Rev.
\textbf{147}, 295 (1966).

\bibitem{Bean} C. P. Bean, Rev. Mod. Phys. \textbf{36}, 31 (1964).

\bibitem{Malozemoff} A. P. Malozemoff, Physica C \textbf{185-189}, 264
(1991).

\bibitem{Fisher} M. P. A. Fisher, Phys. Rev. Lett. \textbf{62}, 1415
(1989); D. S. Fisher, M. P. A. Fisher, and D. A. Huse, Phys. Rev. B
\textbf{43}, 130 (1991).

\bibitem{Vinokur} M. V. Feigel'man, V. B. Geshkenbein, A. I. Larkin, and V. M. Vinokur, Phys. Rev. Lett.\textbf{63}, 2303 (1989).

\bibitem{DMRM1} H. G. Schnack, R. Griessen, J. G. Lensink, C. J. van der Beek , and P. H. Kes, Physica C \textbf{197}, 337 (1992).

\bibitem{DMRM2} M. Jirsa, L. Pust, H. G. Schnack, and R.
Griessen, Physica C \textbf{207}, 85 (1993).

\bibitem{YBCOWen} H. H. Wen, H. G. Schnack, R. Griessen, B. Dam, and J. Rector, Physica C \textbf{241}, 353 (1995).

\bibitem{MgB2Wen} H. H. Wen, S. L. Li, Z. W. Zhao, H. Jin, Y. M. Ni, Z. A. Ren, G. C. Che, and Z. X. Zhao, Physica C \textbf{363}, 170 (2001).

\bibitem{YBCO_cp} A. P. Malozemoff and M. P. A. Fisher, Phys. Rev. B
\textbf{42}, 6784(R) (1990).

\bibitem{Griessen} R. Griessen, H. H. Wen, A. J. J. van Dalen, B. Dam, J. Rector,
 H. G. Schnack, S. Libbrecht, E. Osquiguil, and Y. Bruynseraede, Phys. Rev. Lett.
\textbf{72}, 1910 (1994).

\bibitem{Tl_Wen} H. H. Wen, A. F. Th. Hoekstra, R. Griessen, S. L. Yan, L. Fang,
 and M. S. Si, Phys. Rev. Lett. \textbf{79}, 1559 (1997).

\bibitem{JinH} H. Jin, H. H. Wen, H. P. Yang, Z. Y. Liu, Z. A. Ren, G. C. Che, and Z. X. Zhao, Appl. Phys. Lett.
\textbf{83}, 2626 (2003).

\bibitem{Maley} M. P. Maley, J. O. Willis, H. Lessure, and M. E. McHenry, Phys. Rev. B
\textbf{42}, 2639 (1990).

\bibitem{Yamamoto} A. Yamamoto, A. A. Polyanskii, J. Jiang, F.
Kametani, C. Tarantini, F. Hunte, J. Jaroszynski, E. E. Hellstrom,
P. J. Lee, A. Gurevich, D. C. Larbalestier, Z. A. Ren, J. Yang, X.
L. Dong, W. Lu, and Z. X. Zhao, Supercond. Sci. Technol.
\textbf{21}, 095008 (2008).

\bibitem{Prozorov} R. Prozorov, M. E. Tillman, E. D. Mun, and P. C. Canfield, arXiv:0805.2783.

\bibitem{Dou} X. L. Wang, S. R. Ghorbani, S. X. Dou, X. L. Shen, W. Yi, Z. C. Li, Z. A.
Ren, arXiv:0806.1318.

\end{thebibliography}
\end{document}